# Advancements in Orthopaedic Arm Segmentation: A Comprehensive Review

Abhishek Swami, Snehal Farande,
Atharv Patil, Atharva Parle, Vivekanand Mane, Prathamesh Thorat

Department of Computer Science & Engineering (AIML)
**D. Y. Patil College of Engineering and Technology, Kolhapur**

___

*Abstract — The most recent advances in medical imaging that have transformed diagnosis, especially in the case of interpreting X-ray images, are actively involved in the healthcare sector. The advent of digital image processing technology and the implementation of deep learning models such as Convolutional Neural Networks (CNNs) have made the analysis of X-rays much more accurate and efficient. In this article, some essential techniques such as edge detection, region-growing technique, and thresholding approach, and the deep learning models such as variants of YOLOv8 – which is the best object detection and segmentation framework – are reviewed. We further investigate that the traditional image processing techniques like segmentation are very much simple and provides the alternative to the advanced methods as well. Our review gives useful knowledge on the practical usage of the innovative and traditional approaches of manual X-ray interpretation. The discovered information will help professionals and researchers to gain more profound knowledge in digital interpretation techniques in medical imaging.*

## INTRODUCTION

The introduction of image segmentation has been a huge game changer between computer vision and machine learning and as much in the field of X-ray screenings in medical diagnosis. This technique has emerged as irreplaceable when making a diagnosis and planning the treatment [1]. This modern-day technology may be a real game-changer in terms of administering medical decision-making procedures with more precision and accuracy. With its pixel-by-pixel feature, image segmentation is capable of isolating structures in radiological images, such as bones and tissues with little or no anomalies. Thus, image segmentation is considered as a highly efficient tool with a number of applications. The members of medical cared observed the bone mending prowess in the detection of abnormalities, bone fracture and disease diagnosis.

Traditional approaches for medical image segmentation utilised by professionals or by hand-crafted designs based on image processing operations have been thresholding [2], edge detection [3], and morphological operations. These ways enable to a degree of the interpretability and control, which are useful in some applications. The classic approaches to the segmentation task, however, have obvious limitations because the images and the segmentation situations in medical imaging are complicated and diverse. Customised algorithms can not keep up with processing speed and low accuracy when it comes to numerous scans. In addition to this, the manual feature extraction from medical images requires experts who have the expertise and experience. So, their task is prone to humans' uncertainties.

In recent years, deep learning algorithms have been widely used in medical image segmentation to overcome these concerns. Deep feature learning allows models to extract semantic information from images, improving segmentation accuracy and allowing them to adapt to different medical image datasets and tasks. Since convolutional neural network (CNN)-based segmentation models have achieved remarkable results, the combination of image segmentation and X-ray diagnosis has evolved in such a way that it not only increases the precision of medical evaluations but also equips medical professionals to design more informed and clever treatment plans, improving patient outcomes. However, despite these advances, there are still difficulties associated with interpreting X-ray images for various diseases, considering the variations in patient anatomy, and understanding different bone structures. Despite these obstacles, image segmentation has the revolutionary potential to redefine medical diagnosis and healthcare decision-making. The basic idea of automating the X-ray diagnosis procedure highlights the path towards a more sophisticated, efficient, and significant medical future.

We will review the method for segmenting Regions of Interest (ROIs) in X-ray Images, primarily using the YOLOv8 model [4]. We aim to take advantage of the strengths of Convolutional Neural Network (CNN) architectures, specifically YOLOv8, to achieve precise and efficient segmentation. The YOLOv8 model, known for its exceptional object detection capabilities, is the foundation of our method. It allows us to precisely identify and localise ROIs across a wide range of medical imaging, improving the accuracy and efficiency of the segmentation process [5].

# OVERVIEW OF TRADITIONAL IMAGE SEGMENTATION TECHNIQUES

Traditional image processing techniques for segmentation are essential techniques used to identify objects of interest within images [6]. Following are some traditional segmentation techniques-

1. Thresholding: It is a simple image segmentation method where pixels in an image are categorized into two classes based on whether their intensity values are above or below a specified threshold. Mathematically, it can be formulated as follows: Let $I(x, y)$ represent the intensity value of a pixel at coordinates $(x, y)$. The thresholding operation can be represented as:

$$T(x, y) = \begin{cases} 1, & \text{if } I(x, y) > T_{\text{threshold}} \\ 0, & \text{otherwise} \end{cases}$$

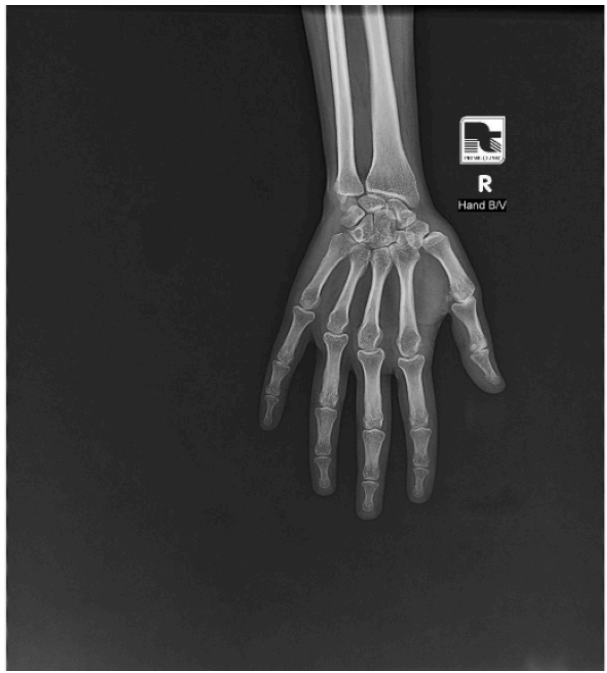
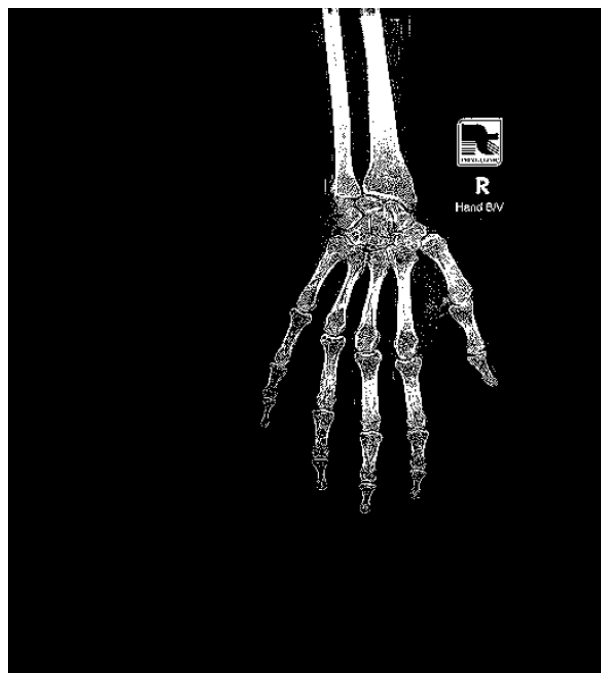

**(a)** Original X-ray image     **(b)** Image formed after thresholding

**Figure 1. Result of segmentation using thresholding.**
Hand X-ray image of size 1280 × 1280 pixels passed through thresholding model with threshold of 177

- Advantages:
    i. Simple and computationally efficient.
    ii. Easy to implement.
- Disadvantages:
    i. Sensitive to noise and variations in pixel intensity.
    ii. May require manual selection of threshold value, which can be subjective.

2. Region Growing: It is a method where similar adjacent pixels are grouped together to form a region based on predefined criteria such as intensity or texture similarity [7]. Mathematically, it can be formulated as follows: Let $I(x, y)$ represent the intensity value of a pixel at coordinates $x$ and $y$ in the image. Let $R$ represent the region being grown, initially containing the seed point $S$. The region growing process can be represented as an iterative algorithm:

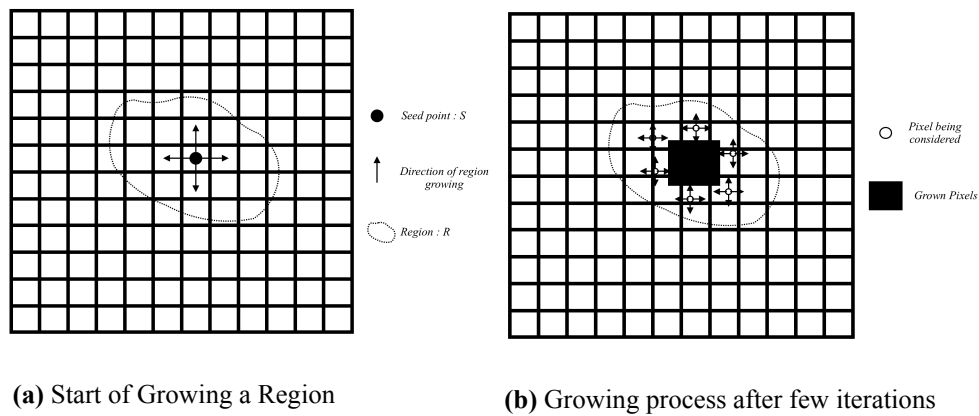

(a) Start of Growing a Region  (b) Growing process after few iterations

**Figure 2. Working of Region Growing Algorithm**

I. Initialize $R$ with the seed point $S$ as shown in Figure 2(a).
II. For each pixel $P$ in the region $R$:
    a. Find neighbouring pixels of $P$.
    b. Calculate the similarity between the intensity of each neighbouring pixel and the average intensity of the pixels in $R$.
    c. If the intensity of a neighbouring pixel is similar to the average intensity of $R$ (based on a predefined threshold), add that pixel to $R$.
III. Repeat step 2 until no more pixels can be added to $R$ or until a stopping criterion is met.

- Advantages:
    i. Can handle irregularly shaped objects and regions.
    ii. Automatically adapts to local image characteristics.
- Disadvantages:
    i. Sensitive to seed point selection.
    ii. Computationally expensive for large images or complex scenes.

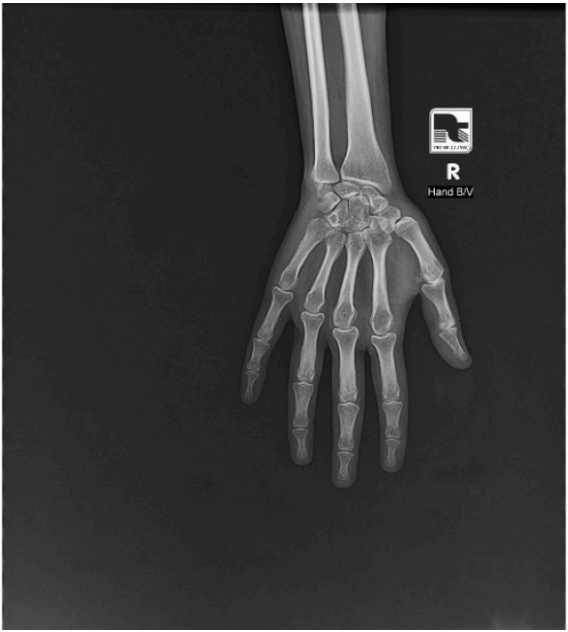 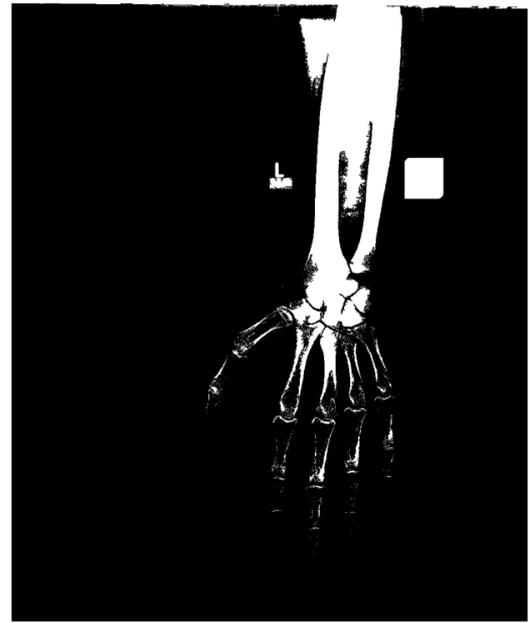

    **(a)** Original X-ray image        **(b)** Image formed using region growing technique

**Figure 3. Result of segmentation using region growing model**
Hand X-ray image of size 1280 × 1280 pixels passed through region growing model at seed point (640,790) with threshold of 60

3. Edge-based Detection: There are several methods in edge-based detection which aim to identify edges or boundaries in an image by detecting discontinuities in intensity or texture. Various edge detection operators can be used, such as Sobel, Prewitt, Roberts, and Canny. These operators typically involve convolving the image with a kernel to compute gradients or edge strengths.

    3.1. Sobel Operator: It is commonly used for edge detection. It calculates the gradient of the image intensity at each pixel, emphasizing edges where there is a significant change in intensity [8]. It consists of two 3x3 convolution kernels, one for horizontal changes and the other for vertical changes:

$$G_x = \begin{bmatrix} -1 & 0 & 1 \\ -2 & 0 & 2 \\ -1 & 0 & 1 \end{bmatrix} \quad G_y = \begin{bmatrix} -1 & -2 & -1 \\ 0 & 0 & 0 \\ 1 & 2 & 1 \end{bmatrix}$$

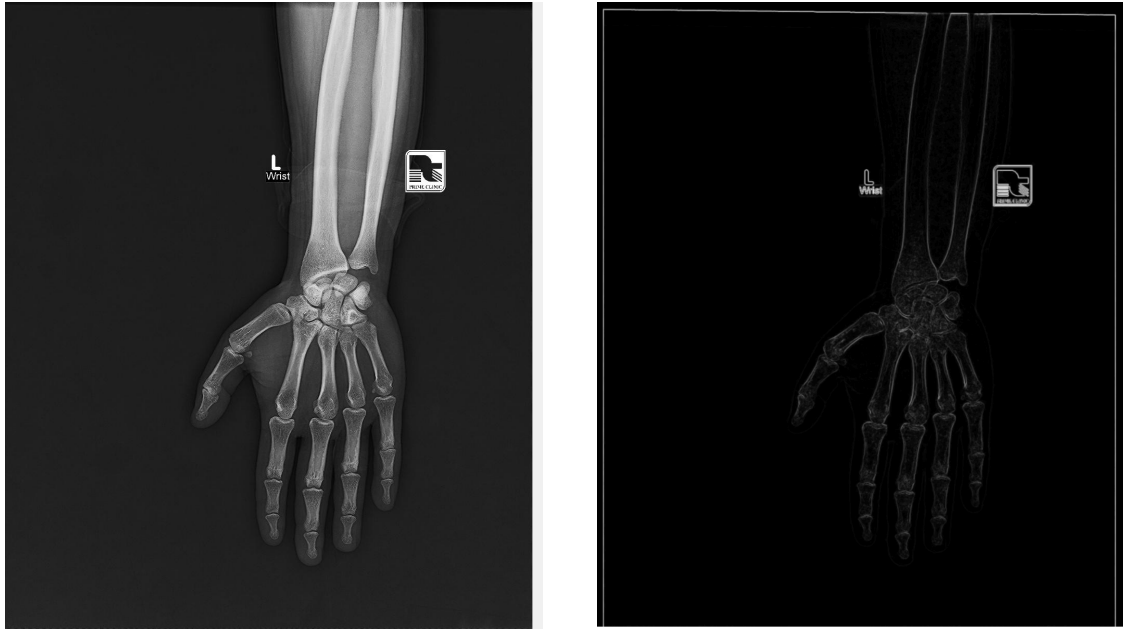

**(a)** Original X-ray image        **(b)** Image formed after applying Sobel Operator

**Figure 4. Result of Edge-based Detection using Sobel Operator**
Hand X-ray image of size 1280 × 1280 pixels operated with Sobel operator (kernel) of size 3 × 3

3.2. Prewitt Operator: It is similar to Sobel operator and is used for edge detection. It calculates the gradient of the image intensity at each pixel in the horizontal and vertical directions [9]. It consists of two 3x3 convolution kernels:

$$G_x = \begin{bmatrix} -1 & 0 & 1 \\ -1 & 0 & 1 \\ -1 & 0 & 1 \end{bmatrix} \quad G_y = \begin{bmatrix} -1 & -1 & -1 \\ 0 & 0 & 0 \\ 1 & 1 & 1 \end{bmatrix}$$

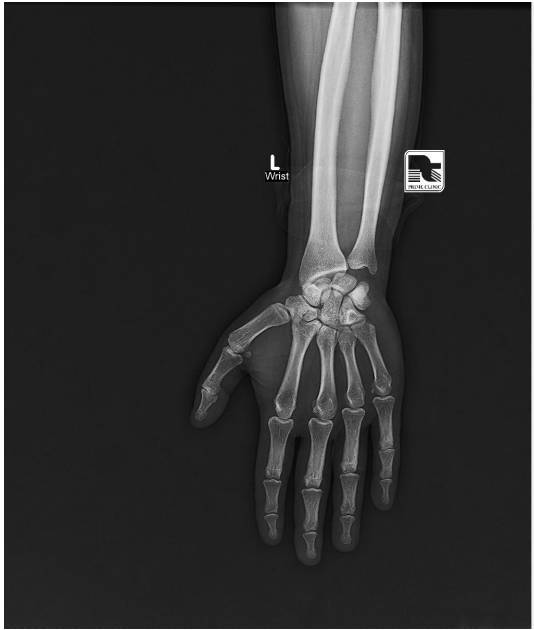 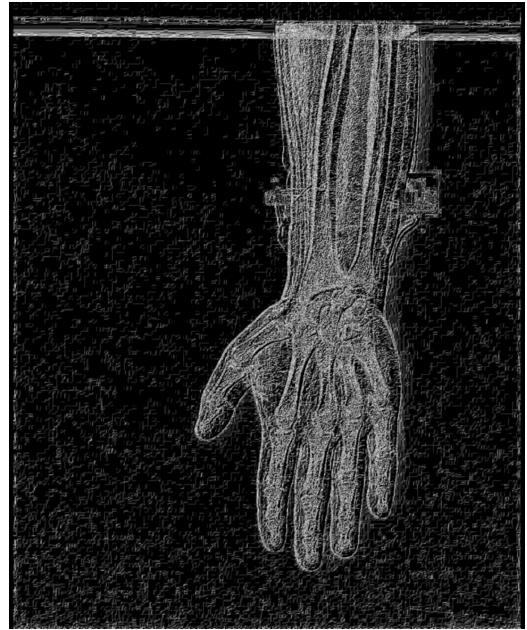

| **(a)** Original X-ray image | **(b)** Edge mask formed after applying Prewitt operator |

**Figure 5. Result of Edge-based Detection using Prewitt Operator**
Hand X-ray image of size 1280 × 1280 pixels operated with Prewitt operator (kernel) of size 3 × 3

3.3. Roberts Operator: It is a simple edge detection operator that calculates the gradient of the image intensity using 2x2 convolution kernels [10]. It consists of two 2x2 convolution kernels:

$$G_x = \begin{bmatrix} 1 & 0 \\ 0 & -1 \end{bmatrix} \quad G_y = \begin{bmatrix} 0 & 1 \\ -1 & 0 \end{bmatrix}$$

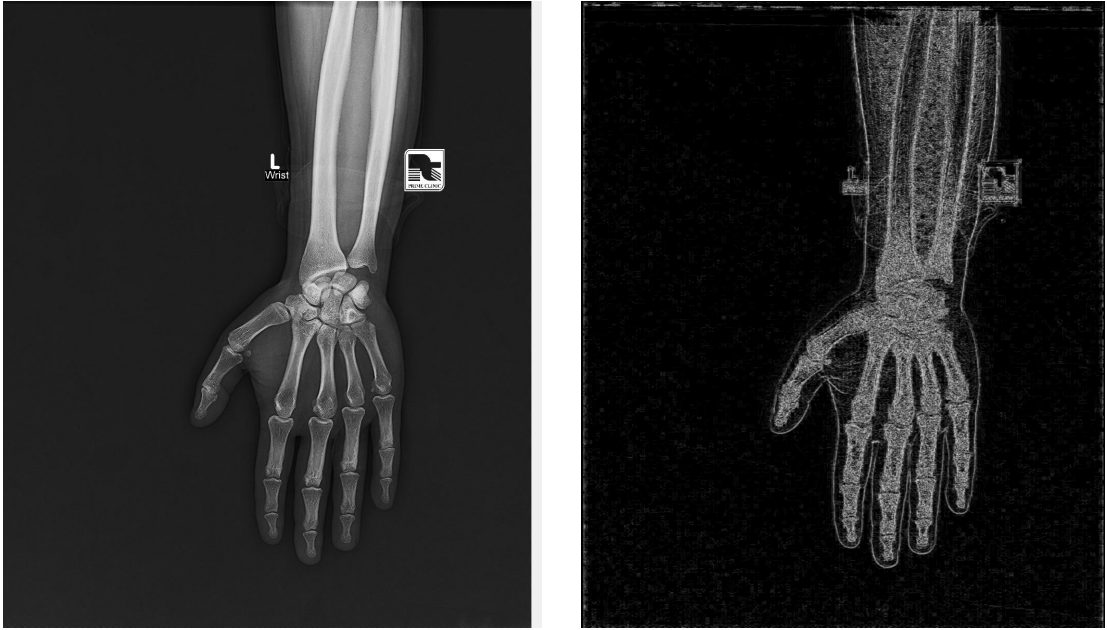

|       (a) Original X-ray image       | (b) Edge mask formed after applying Roberts Operator |

**Figure 6. Result of Edge-based Detection using Roberts Operator**
Hand X-ray image of size 1280 × 1280 pixels operated with Roberts operator (kernel) of size 2 × 2

3.4. Canny Operator: It is a multi-stage algorithm used for edge detection, which includes smoothing, gradient calculation, non-maximum suppression, and edge tracking by hysteresis [11]. The specific kernel matrices used in Canny operator are not fixed as they depend on the Gaussian smoothing and gradient calculation steps, which are usually performed using convolution. These are the matrices for some commonly used edge detection operators. These operators are applied to the image using convolution to detect edges or boundaries.

- Advantages:
    i. Can provide precise delineation of object boundaries.
    ii. Robust to noise.
- Disadvantages:
    i. May produce multiple edges in complex scenes.
    ii. Prone to variations in intensity and noise.

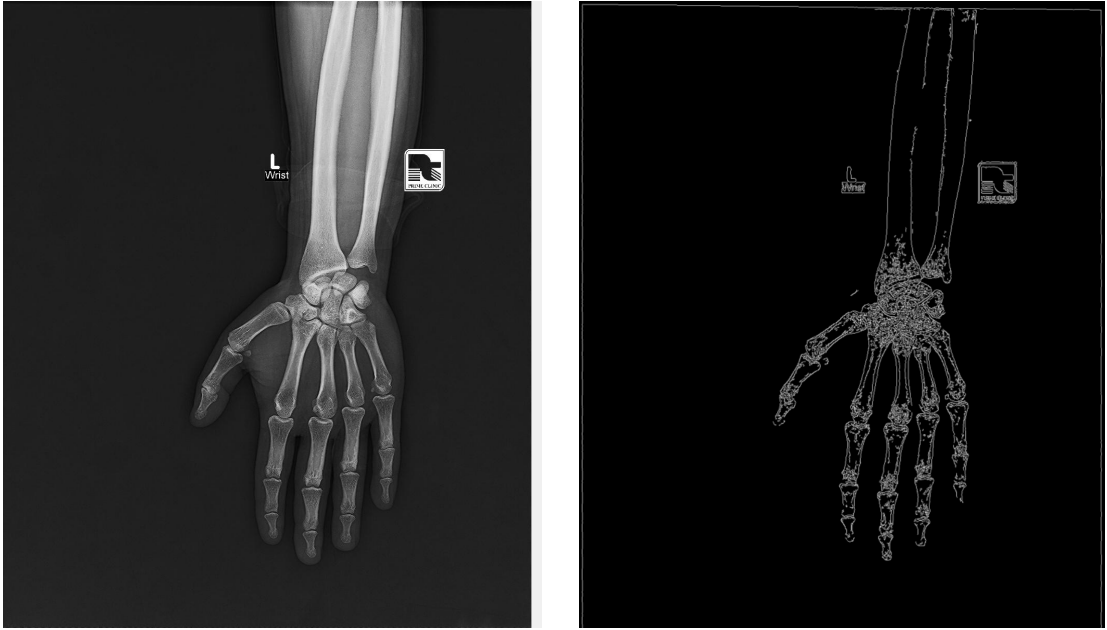

**(a)** Original X-ray image        **(b)** Edge detected using Canny Operator

**Figure 7. Result of Edge-based Detection using Canny Operator**

4. Active Contour Model (Snake): They are deformable models that evolve to fit the boundaries of objects in an image based on energy minimization principles. Active contour models are represented by a curve $C(s)$, where $s$ represents the arc-length parameter [12]. The energy functional $E(C)$ is minimized iteratively to deform the contour towards object boundaries.

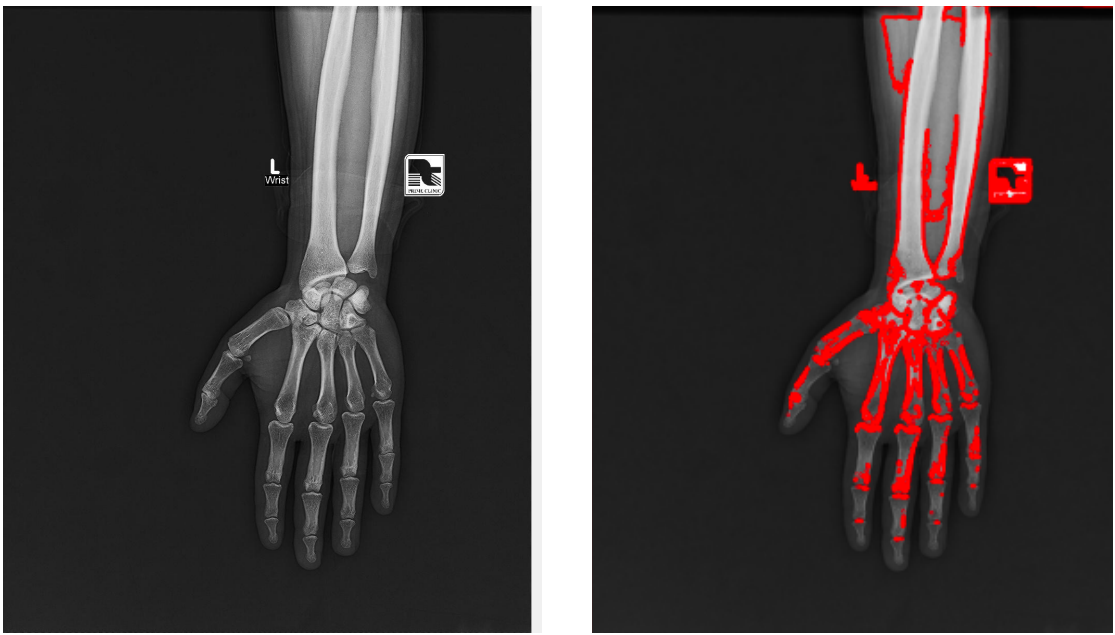

**(a)** Original X-ray image        **(b)** Edge detection using active contour model

**Figure 8. Result of Edge-based Detection using Active Contour Model (Snakes)**

- Advantages:
    i. Can handle complex shapes and irregular boundaries.
    ii. Robust to noise and occlusions.
- Disadvantages:
    i. Sensitive to initialization and parameter tuning.
    ii. Computationally intensive, especially for large images or complex objects.

These are basic explanations of each segmentation method along with their mathematical formulations, results, advantages, and disadvantages. Further details and optimizations can be explored based on specific applications and requirements. Thresholding stands out as a basic yet efficient technique among these. It differentiates objects from the background by applying an intensity threshold, classifying pixels with intensities above the threshold as part of the object and those with values below the threshold as background. Moreover, morphological treatments such as erosion and dilation change the shape and structure of segmented objects [13].

These methods, which are frequently used for post-processing, can refine segmentation results and improve object connectivity. While these classic methods have their advantages, they might fail when faced with complicated, noisy, or highly changeable images. These methods have significant constraints in the context of X-ray images, which may limit their usefulness. X-ray scans frequently show different levels of contrast, intricate anatomical features, and the possibility of abnormalities. Because the intensity differences between objects and background can be insignificant, thresholding techniques that require pixel intensity values might fail to give correct segmentations in such scenarios. Besides, complex shapes and boundaries, which are often seen in X-ray images, may offer difficulties for edge detection systems. Additionally, the sensitivity of X-rays to image noise might result in the inclusion of undesired artefacts, decreasing the accuracy of the segmentation results. Traditional techniques may also include manual parameter adjustments, making them impractical for the quick processing of huge quantities of X-ray images.

**YOLOv8 AND ITS SIGNIFICANCE**

YOLOv8 is a state-of-the-art object detection and image segmentation model, representing the latest iteration of the YOLO series, which stands for "You Only Look Once" [14]. It has achieved significant performance in computer vision and deep learning.

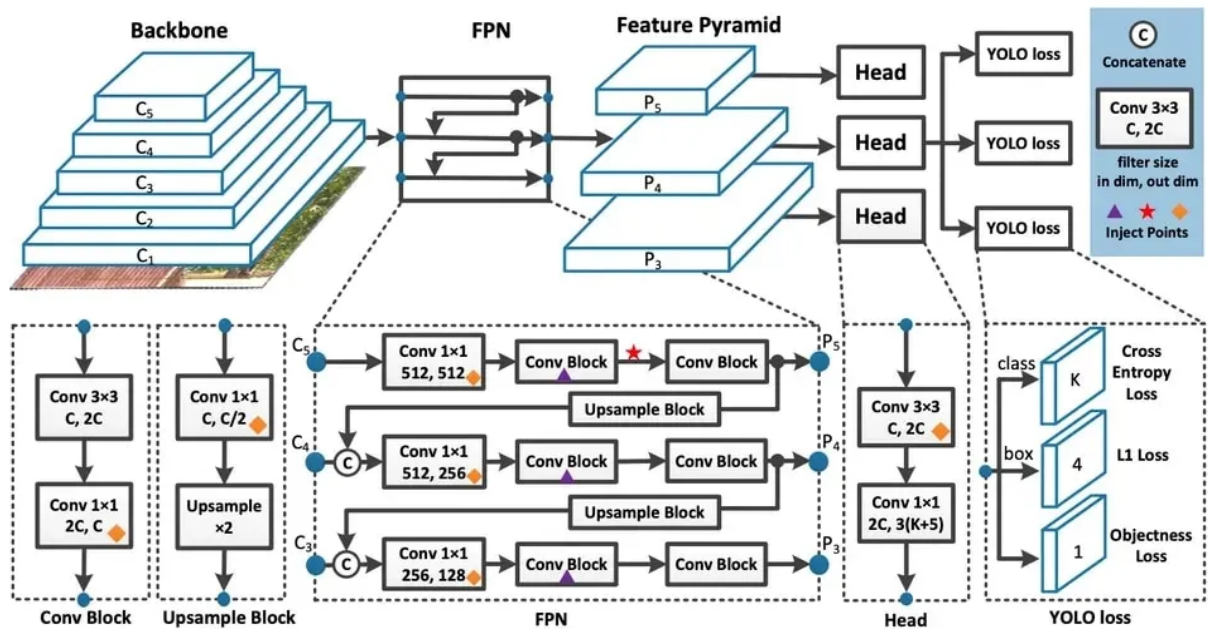

Figure 9. YOLOv8 Architecture

Here is an overview of YOLOv8 architecture:

- Backbone Network: YOLOv8 typically uses a powerful backbone convolutional neural network (CNN) such as DarkNet or ResNet as its base. This backbone network is responsible for extracting features from the input image at multiple scales and levels of abstraction.

- Feature Pyramid Network (FPN): YOLOv8 often incorporates a feature pyramid network (FPN) or similar architecture to capture multi-scale features from different levels of the backbone network. FPN helps in detecting objects of varying sizes and scales by combining features from different layers.

- Detection Head: YOLOv8 employs a detection head or prediction module that predicts bounding boxes, objectness scores, and class probabilities for the objects present in the image. The detection head typically consists of a series of convolutional and fully connected layers that process the features extracted by the backbone network.

- Anchor Boxes: YOLOv8 uses anchor boxes or priors to predict bounding boxes for objects. These anchor boxes are predefined shapes with different aspect ratios and scales, which are used as references for predicting the bounding box coordinates.

- Output Format: YOLOv8 outputs detections in a grid-based format. Each grid cell predicts multiple bounding boxes along with confidence scores for object presence and class probabilities. The final detections are obtained by applying non-maximum suppression (NMS) to filter out redundant and overlapping bounding boxes.

- Loss Function: YOLOv8 typically uses a combination of loss functions, including localization loss, confidence loss, and classification loss, to train the network. These loss functions penalize errors in bounding box localization, objectness prediction, and class prediction, respectively.

**Significance:**

- Real-Time Object Detection: YOLOv8 is known for its real-time object detection capabilities, allowing it to recognise and locate objects within images or video frames efficiently. This real-time capability is necessary for applications like surveillance, autonomous vehicles, and augmented reality, where fast decision-making based on object detection is essential [15].

- Single Forward Pass: YOLO's "You Only Look Once" concept tells that it performs object detection and localisation in a single forward pass through the neural network. This efficiency is the opposite of other object detection methods that require multiple passes, making YOLOv8 faster and more computationally efficient.

- Versatility: Versatility is a significant aspect of YOLOv8, making it suitable for various object detection tasks, including detecting objects in natural scenes, tracking objects over time, and even segmenting objects.

- Accurate Object Localisation: YOLOv8 provides precise object localisation, meaning it not only detects objects but also accurately outlines their boundaries. This high degree of accuracy is crucial in applications where knowing the exact object location is essential, such as in medical image analysis or autonomous navigation.

- Open Source and Community Support: YOLOv8, like its older versions, is open-source, making it accessible to the researchers and developer community. This open nature encourages collaboration, improvement, and the development of diverse applications.

- Transfer Learning: YOLOv8 can be fine-tuned and trained for specific tasks through transfer learning, which enables researchers and developers to build custom object detection models with less data and effort.

- State-of-the-Art Performance: YOLOv8 has pushed the boundaries of object detection performance. It consistently achieves high mean average precision (mAP) scores on benchmark datasets, indicating its state-of-the-art accuracy.

- Practical Applications: YOLOv8's combination of speed and accuracy has led to its use in various practical applications, including security and surveillance, self-driving cars, robotics, medical image analysis, and more. It has the potential to improve the efficiency and safety of various industries.

YOLOv8's significance lies in its ability to provide real-time, accurate, and versatile object detection capabilities, making it a valuable tool for multiple applications in computer vision and beyond. Its efficiency and open-source nature have made it a popular choice for researchers and developers.

# COMPARISON OF YOLOv8 VARIANTS

Real-time object detection has attracted a great deal of interest in the development of the YOLO (You Only Look Once) series that balances precision and swiftness. It is the latest instalment from which innovations have been made to offer adaptability for multiple object detection tasks over its predecessors [16].

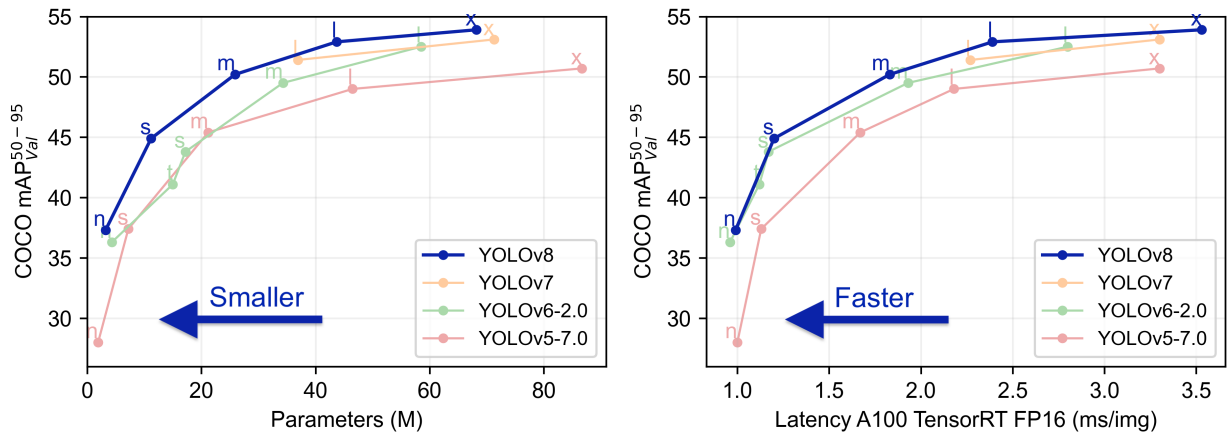

**Figure 10. YOLO models comparison plots**

This analysis compares different YOLOv8 variations based on performance metrics with a focus on how effective they are in segmentation tasks. The comparison includes model size, mAP for bounding boxes and masks, inference speeds on different hardware platforms as well as model complexity measured by parameters and FLOPs. Thus, it is possible to understand what kind of trade-off there is between the size of the model, accuracy and computational efficiency; this can be helpful to researchers or practitioners in choosing an appropriate version for their specific tasks.

Table 1 - Comparison of YOLO-segmentation models of different sizes

| Model | size (pixels) | mAPbox (ms) | MAP mask (ms) | Speed CPU ONNX (ms) | Speed A100 TensorRT (ms) | params (M) | FLOPS (B) |
|---|---|---|---|---|---|---|---|
| **YOLOv8n-seg** | 640 | 36.7 | 30.5 | 96.1 | 1.21 | 3.4 | 12.6 |
| **YOLOv8s-seg** | 640 | 44.6 | 36.8 | 155.7 | 1.47 | 11.8 | 42.6 |
| **YOLOv8m-seg** | 640 | 49.9 | 40.8 | 317 | 2.18 | 27.3 | 110.2 |
| **YOLOv8l-seg** | 640 | 52.3 | 42.6 | 572.4 | 2.79 | 46 | 220.5 |
| **YOLOv8x-seg** | 640 | 53.4 | 43.4 | 712.1 | 4.02 | 71.8 | 344.1 |

The presented data provides a complete comparison of YOLOv8 versions optimized for segmentation tasks. Distinctions between each variant make use of its model size in the form of pixels counting the width and height of input images. The mAP scores concerning bounding boxes as well as masks capture how well the detection of objects is done and their segmentation respectively with higher values indicating more efficient performances. Furthermore, times taken to run inference on CPU and GPU that are ONNX runtime and A100 TensorRT were analyzed to show the real-time capabilities that these models have. Meanwhile, parameters count and FLOPs provide insights into computational complexity for each version–this helps to guide choice-making based on computation resources available as well as application requirements.

**APPLICATION IN X-RAY IMAGE ANALYSIS**

We are focusing on the analysis of X-ray images of arm bones, specifically the radius, ulna, carpals, metacarpals and phalanges. This targeted study starts with the collection and structuring of a specialized dataset comprising X-ray images of these specific arm bone structures. Making accurate masks and labels to highlight and identify different bones in these photos is an essential component of this process [17]. By generating labels in a format suitable for YOLOv8, which is the ".txt" file format, we aim to enable the model to recognize and distinguish these bones with accuracy, simplifying the diagnosis of X-ray images.

1. Dataset Preparation:
   - Data Collection: Collecting X-ray images that feature arm bones. These images should represent various arm bone structures, such as the radius, ulna, carpals, metacarpals and phalanges.
   - Data Organization: Structuring dataset in a way recommended for training YOLOv8 model with enhanced and resized X-ray images.

2. Mask Creation and Bone annotation:
   - Manual Annotation: Skilled annotators manually create masks by outlining and labelling each specific arm bone part within the X-ray images, using specific annotator tool.

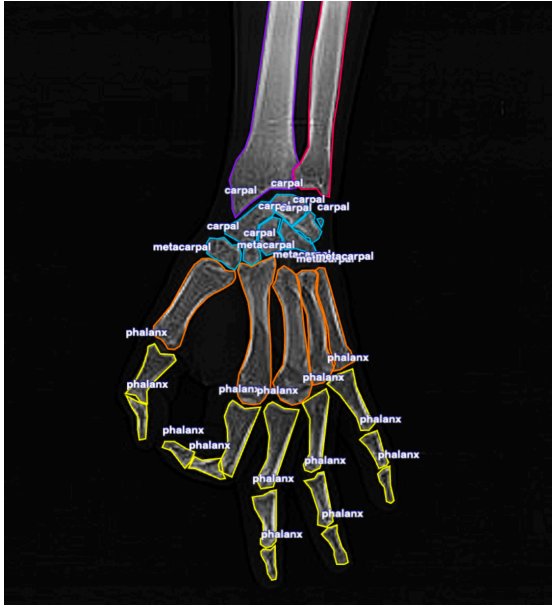 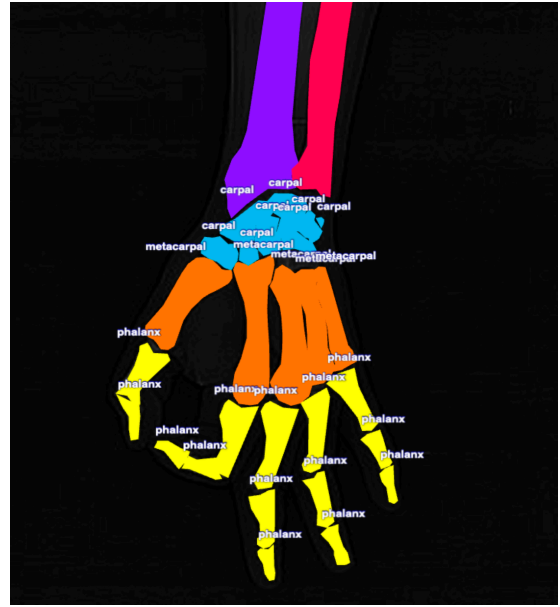

**(a)** An X-ray image of a hand annotated with instance labels using a polygon tool

**(b)** An X-ray image of a hand with layered mask

**Figure 11. Mask Creation and Bone Annotation**

- Bone Identification: Each segmented bone, such as the radius or ulna, is accurately labelled with its name to denote its specific identity.

3. Label generation in YOLOv8 format:
   - Label File Generation: This involves generating YOLOv8-compatible labels for bones. These labels contain information about the class index, object coordinates, and dimensions.

With these steps, a dataset focused on X-ray images of bones, specifically structured for YOLOv8 segmentation can be prepared. This streamlined approach showcases the analysis of X-ray images with higher accuracy, which can be helpful for the precise diagnosis of underlying conditions.

# RESULT AND ANALYSIS

We assessed how well the segmentation model can identify and outline different parts of bones. To do this, we used facility in YOLOv8 base model provided by ultralytics and the Weights&Biases tool (WandB.ai) for the evaluation. The experimental work involved multiple image preprocessing steps, the design of a custom CNN for bone fracture prediction, and the use of YOLOv8 for instance segmentation of bone subparts which helped in the development of RadianceAI, contributing to its enhanced X-ray diagnosis capabilities using machine learning.

- **Model Performance**

The YOLOv8x-seg model with 295 layers, 71726434 parameters, 0 gradients, 343.7 GFLOPs was trained on custom data of X-Ray images of bones such as radius, ulna, carpels, metacarpals and phalanges. Training of this model was carried out on AI server facility provided by college, which consists of 6 NVIDIA RTX A5000 GPUs each with 24256 MB GPU memory. The training process was carried out for 200 epochs, with the best performance achieved on the 93rd epoch. After that, the model did not show significant improvement for 20 epochs and it encountered early stopping.

```
      Epoch    GPU_mem   box_loss   seg_loss   cls_loss   dfl_loss  Instances       Size
    110/200     3.65G      1.246      1.908     0.8357      1.108          2        640: 100%| | 222/222 [00:36<00:00,  6.11it/s]
                 Class     Images  Instances      Box(P          R      mAP50  mAP50-95)     Mask(P          R      mAP50  mAP50-95): 100%| |
                   all         37        759      0.833      0.817      0.866      0.652       0.83      0.815      0.873      0.627

      Epoch    GPU_mem   box_loss   seg_loss   cls_loss   dfl_loss  Instances       Size
    111/200     3.47G      1.173      1.803     0.7722      1.071         22        640: 100%| | 222/222 [00:36<00:00,  6.09it/s]
                 Class     Images  Instances      Box(P          R      mAP50  mAP50-95)     Mask(P          R      mAP50  mAP50-95): 100%| |
                   all         37        759      0.833      0.751      0.853      0.629       0.83      0.748      0.849       0.61

      Epoch    GPU_mem   box_loss   seg_loss   cls_loss   dfl_loss  Instances       Size
    112/200     3.66G      1.122      1.683     0.7021      1.041          8        640: 100%| | 222/222 [00:36<00:00,  6.07it/s]
                 Class     Images  Instances      Box(P          R      mAP50  mAP50-95)     Mask(P          R      mAP50  mAP50-95): 100%| |
                   all         37        759        0.8       0.76      0.849       0.64      0.798      0.758      0.842      0.624

      Epoch    GPU_mem   box_loss   seg_loss   cls_loss   dfl_loss  Instances       Size
    113/200     3.27G      1.172      1.745     0.7868      1.084         76        640: 100%| | 222/222 [00:36<00:00,  6.08it/s]
                 Class     Images  Instances      Box(P          R      mAP50  mAP50-95)     Mask(P          R      mAP50  mAP50-95): 100%| |
                   all         37        759       0.79      0.683      0.797      0.583      0.788      0.682        0.8      0.581
Stopping training early as no improvement observed in last 20 epochs. Best results observed at epoch 93, best model saved as best.pt.
To update EarlyStopping(patience=20) pass a new patience value, i.e. `patience=300` or use `patience=0` to disable EarlyStopping.
```

Figure 12. YOLOv8 Segmentation model training epochs

The model was trained to detect and segment various bones and fractures in input arm X-Ray image, and it successfully achieved this objective. The training time took over 1 hour and 8 minutes.

- **Data Sources**

For our analysis, we collected X-ray images of hands from a diverse range of data repositories and platforms. These included well-known resources like Roboflow and Hugging Face, as well as various government digital libraries.

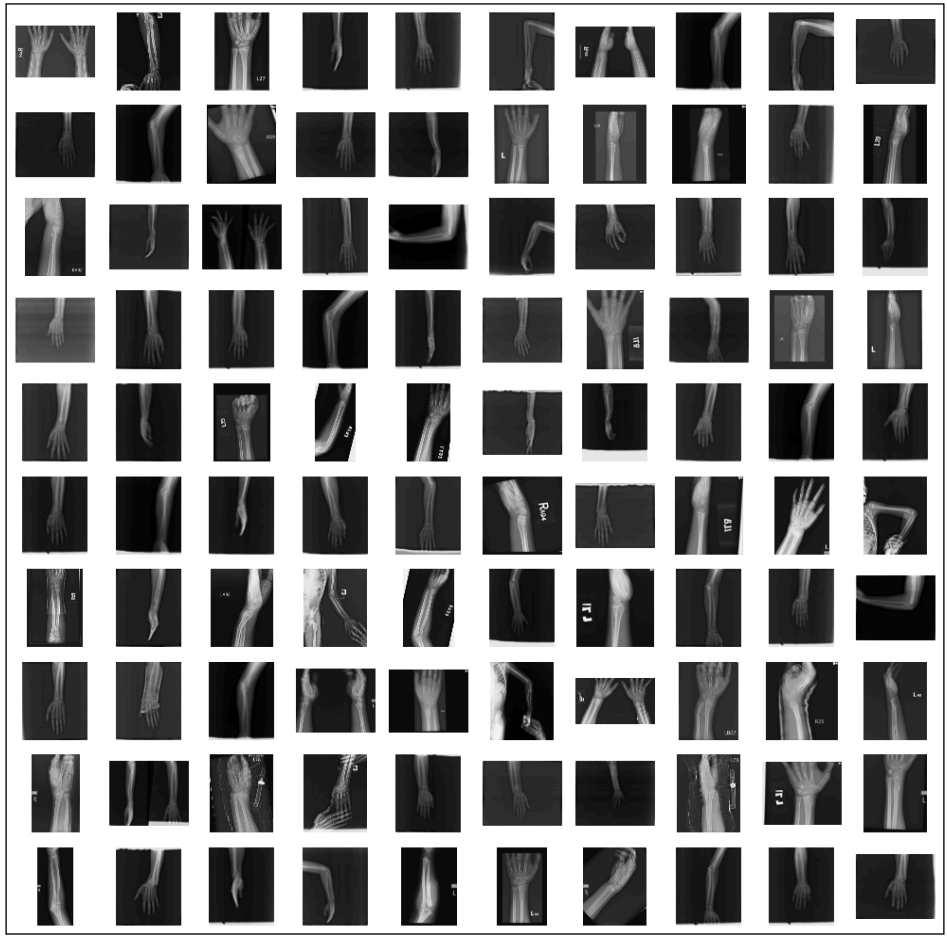

**Figure 13. Raw X-ray Images**

- **Evaluation Metrics**

  - Precision (P): This represents the proportion of correctly identified objects out of all the objects the model identified as that class. A high precision means the model makes few mistakes (low false positives).

  - Recall (R): This represents the proportion of correctly identified objects out of all the actual objects present in the image. A high recall means the model misses few objects (low false negatives).

  - mAP (mean Average Precision): This is a single value that summarises the overall performance across different confidence thresholds. It essentially takes the average of the Precision-Recall Curve (PRC) area for various thresholds. There are two variations mentioned here:

  - mAP50: This refers to the mAP calculated using a specific confidence threshold of 50%.

  - mAP50-95: This refers to the mAP calculated by averaging the PR curve area across multiple confidence thresholds ranging from 50% to 95%.

  The performance of the model based on evaluation set was evaluated using various metrics provided in following table -

| Class | Images | Instances | Box | | | | Mask | | | |
|---|---|---|---|---|---|---|---|---|---|---|
| | | | Precision | Recall | mAP50 | mAP50-95 | Precision | Recall | mAP50 | mAP50-95 |
| all | 37 | 759 | 0.88 | 0.824 | 0.898 | 0.674 | 0.876 | 0.821 | 0.891 | 0.617 |
| carpal | 37 | 204 | 0.862 | 0.673 | 0.865 | 0.581 | 0.85 | 0.665 | 0.837 | 0.472 |
| fracture | 37 | 10 | 0.715 | 0.6 | 0.719 | 0.315 | 0.715 | 0.6 | 0.719 | 0.403 |
| metacarpal | 37 | 163 | 0.905 | 0.94 | 0.96 | 0.801 | 0.905 | 0.94 | 0.96 | 0.718 |
| phalanx | 37 | 318 | 0.885 | 0.872 | 0.907 | 0.713 | 0.872 | 0.86 | 0.894 | 0.543 |
| radius | 37 | 31 | 0.946 | 0.935 | 0.976 | 0.817 | 0.945 | 0.935 | 0.976 | 0.808 |
| ulna | 37 | 33 | 0.968 | 0.925 | 0.96 | 0.814 | 0.968 | 0.925 | 0.96 | 0.759 |

**Table 2 - Evaluation of model based on different metrics**

- **Interpreting the Table:**

  - Class: This indicates the category of object the model is trying to detect (e.g., carpal, fracture).

  - Images: This shows the number of images used for evaluation in this class.

  - Instances: This represents the total number of actual objects present in the images for this class.

  - Box (Metric): These metrics refer to the performance of the model in detecting bounding boxes around the objects.

  - A high value (>0.8) in Box (P) and Box (R) indicates good precision and recall for bounding boxes in that class.

  - mAP50 and mAP50-95 for boxes provide a more comprehensive picture of the model's performance across different confidence levels.

  - Mask (Metric): Similar to bounding boxes, these metrics evaluate the model's performance in predicting segmentation masks around the objects (if applicable).

Looking at the "carpal" class, we see Box (P) is 0.862, indicating the model is good at identifying correct bounding boxes for carpals with relatively few false positives. However, Box (R) is 0.673, suggesting the model might miss some actual carpals (false negatives). The mAP values provide a more complete picture of performance across various confidence levels.

This table provides insights into how well the model performs in detecting different object classes in the dataset. By analyzing precision, recall, and mAP, we can assess the trade-off between correctly identifying objects and missing some.

- **Confusion Matrix :**

    The confusion matrix is a table that summarises the performance of the model by showing true vs predicted results across different bone classes. It helps in understanding how well the model is classifying the objects.

    As we can observe and analyse confusion matrix in Figure.14, True Positive Rate of our model is higher since it is correctly identifying and and predicting mask for segmentation for each bone type. We can also interpret that model can be trained on more data to reduce True Negatives and False Positives.

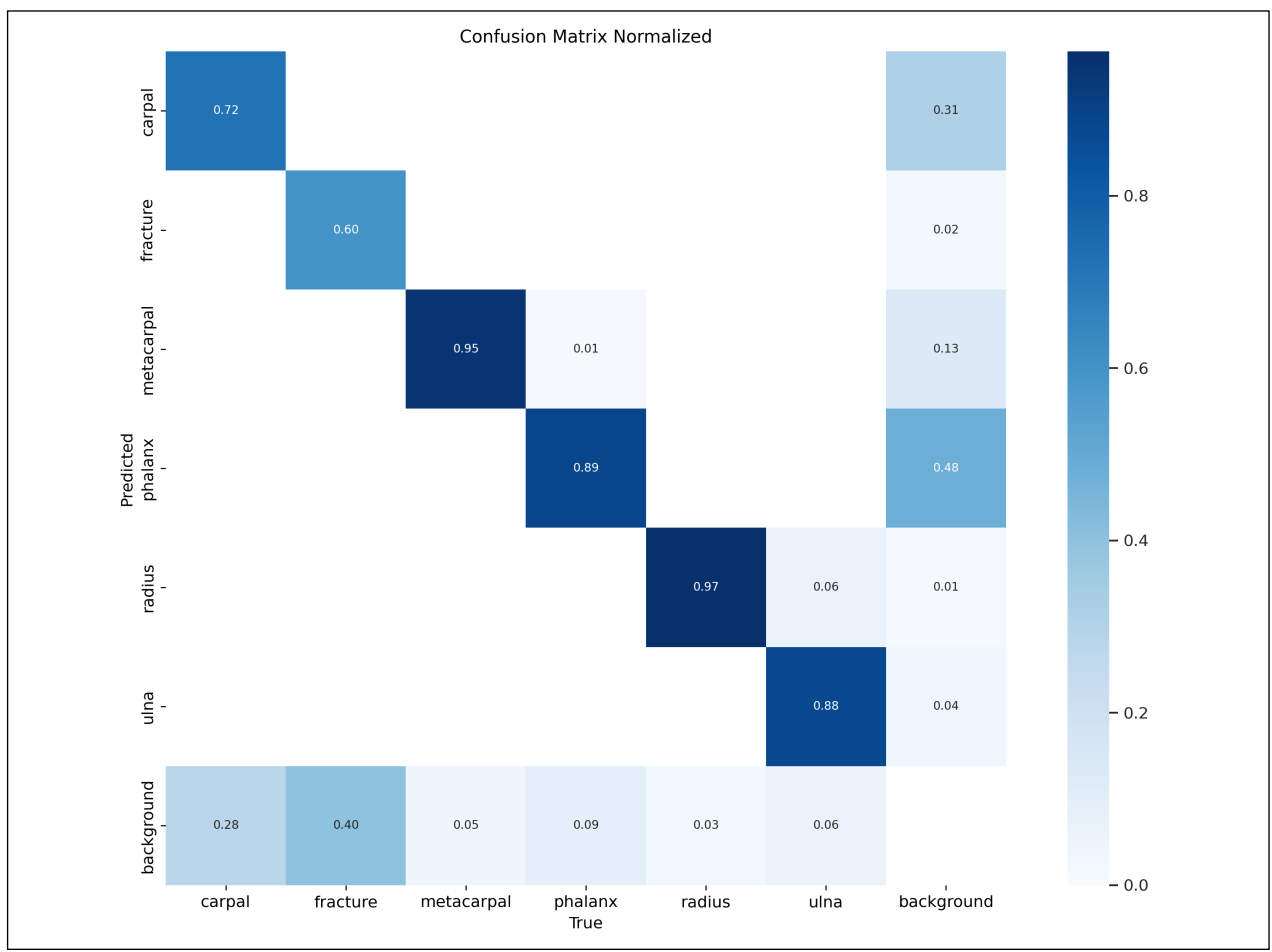

**Figure 14. Confusion Matrix**

- F1 Confidence Curve :

The F1 confidence curve shows the relationship between the F1 score and the confidence threshold. It helps in determining the optimal confidence threshold for the model. As it can be observed from Figure 7.4 , F1-score increases as confidence threshold is increased for all classes and suddenly drops after 0.963.

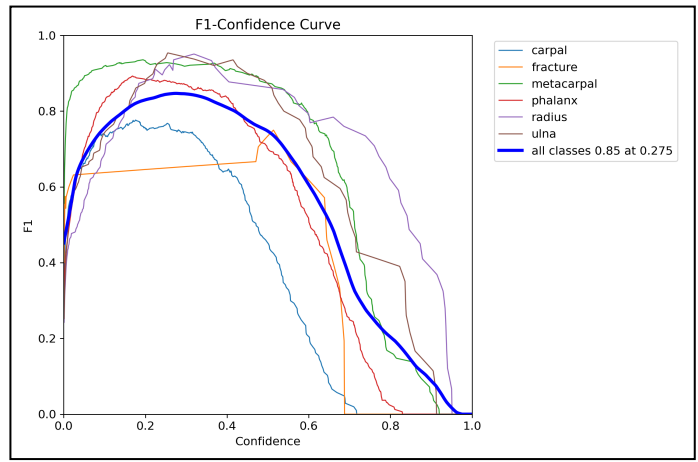

**Figure 15. F1 vs Confidence curve**

- Precision-Confidence Curve :

The precision curve shows the relationship between precision and confidence. In Figure. 7.5, it can be observed that, with confidence threshold, precision of the model increases. From 0.79 to 1.00 confidence threshold, an exponential increase in precision value can be observed.

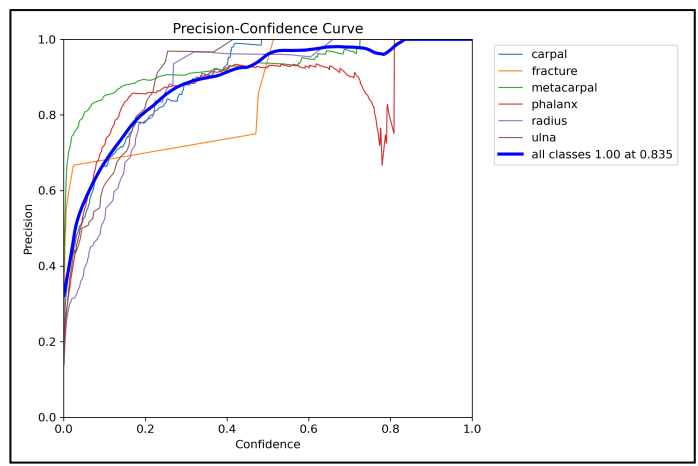

**Figure 16.  Precision vs Confidence curve**

- Precision-Recall Curve :

    The precision-recall curve is a graph that shows the relationship between precision and recall at different confidence thresholds. It helps in determining the optimal confidence threshold for the model. In Figure 7.6, it can be observed that precision and recall are reciprocal of each other at mean average precision i.e. mAP 0.5 for all classes.

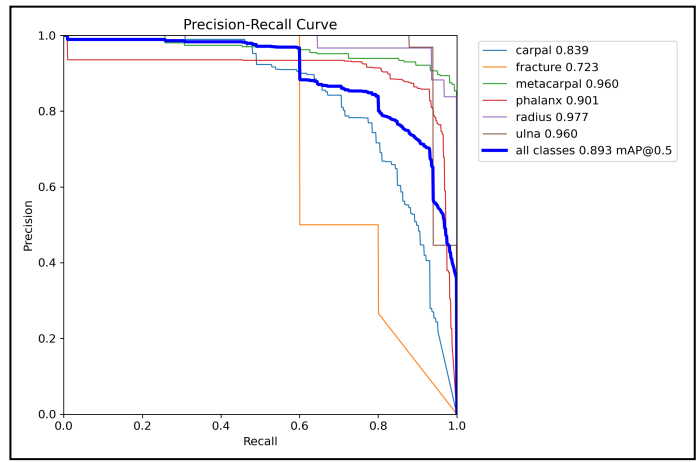

Figure 17. Precision vs Recall curve

- Recall-Confidence Curve :

    Recall confidence is a term commonly used in object detection models, such as YOLO (You Only Look Once). It is a measure of how well the model can detect all the objects in an image, without missing any. Figure 7.7 suggest that, confidence value of 0.6 can deliver better results at optimal recall value.

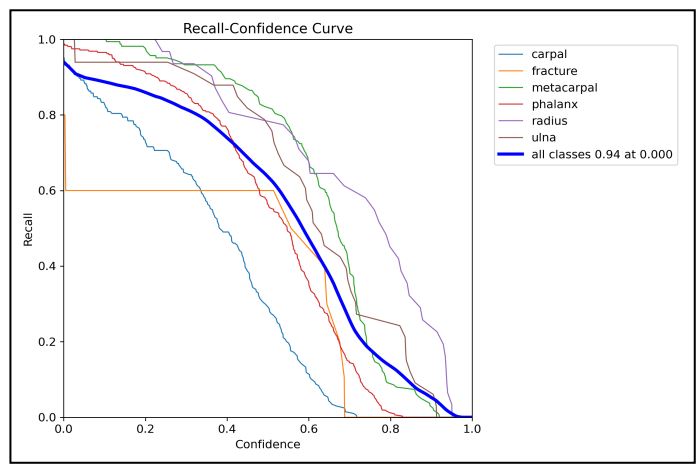

Figure 18. Recall vs Confidence curve

- **Validation Batch Outputs :**

The outputs of training and validation batches were observed to analyse the model's performance and make necessary adjustments. During the training process, the model's outputs were compared with the ground truth labels to compute the loss function. The loss function was used to update the model's parameters to improve its performance. The validation set was used to evaluate the model's performance on unseen data.

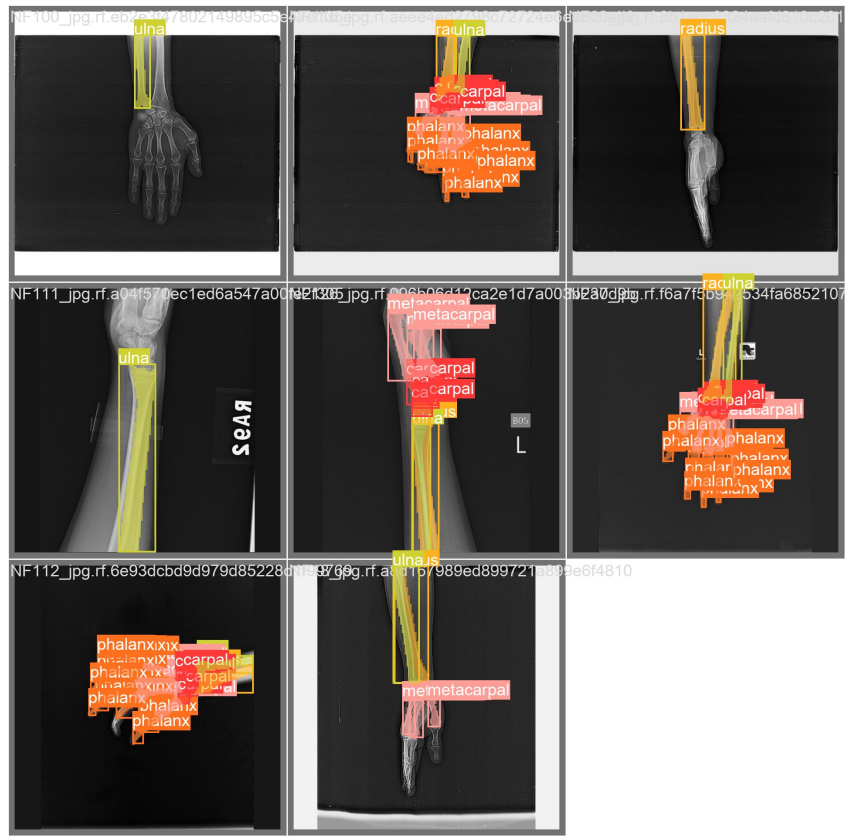

Figure 19. Validation Batch Output

The YOLOv8-seg model trained on the given dataset has shown good performance in detecting and segmenting different bones. The use of diverse images and the evaluation of various metrics helped to assess the model's performance. However, further testing and validation on a larger dataset with varying conditions may be required to confirm its robustness.

- **Inference**

Throughout our work with hand X-ray pictures, we've made a number of essential image enhancements. This includes sharpening the photos to highlight critical features and adjusting brightness and contrast with gamma correction to provide a sharper view of bone architecture.

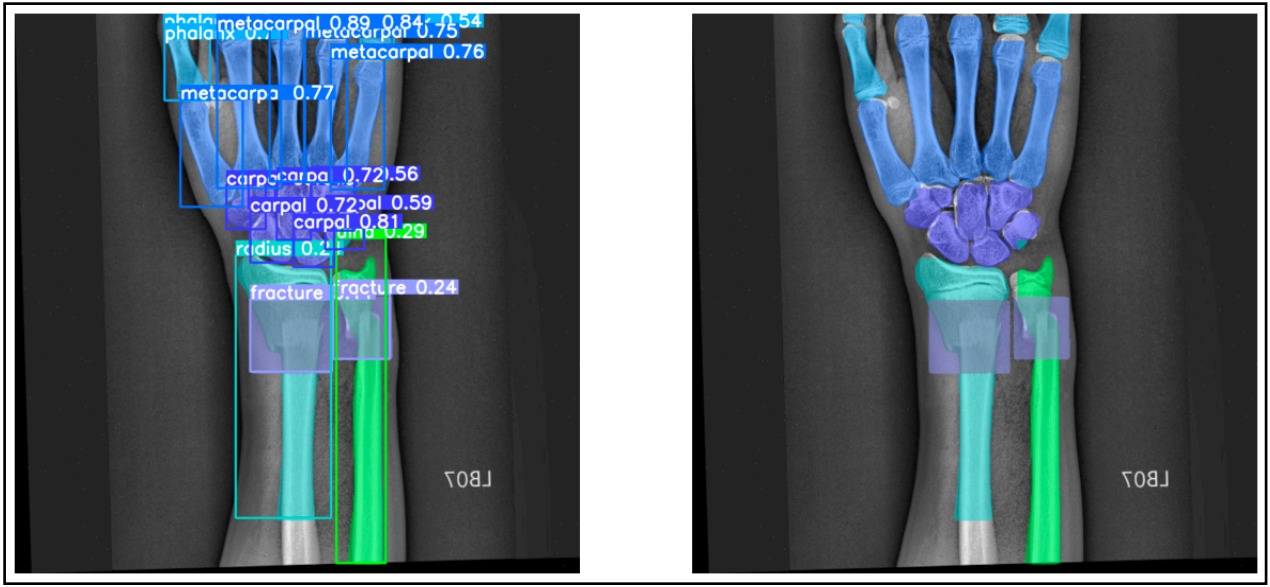

Figure 20. X-ray images of a hand with predicted bone instance segmentation masks and boxes

Edge detection was also used in our investigation, which allowed us to enhance the curves and edges of the hand's bones, making them stand out clearly. We next moved on to binarization, which included converting the photos into a binary representation and separating the foreground (which included the delicate hand anatomy) from the backdrop. These strategic image processing procedures completed the path for accurate segmentation and extensive analysis, eventually advancing medical diagnostics and research.

**CHALLENGES AND LIMITATIONS**

YOLOv8 has shown a state-of-the-art results in object detection and segmentation tasks, but as we consider medical image analysis, we encounter distinct challenges and limitations. Challenges include the intricacy of manual annotation, the variability in human anatomy, noise and artefacts in X-ray images, and the limited data. Also, it operates within certain limitations, such as the model's capacity to generalise, availability of enough computational resources etc.

- Challenges:
1. Precise annotation: Accurate manual annotations of various bones can be difficult due to complex bone structures, which may overlap in X-ray images.
2. Anatomy variability: Human anatomy changes from person to person. Building a model that generalises across various individuals might be a difficult task.
3. Noise and artefacts: X-ray scans often contain noise and artefacts, which can make the segmentation task difficult.
4. Data limitations: Obtaining X-ray images can be tough, especially when working with specific types of bones. Limited data may affect the model's ability to generalise effectively.

- Limitations:
1. Generalisation: The model may have difficulty generalising new, unseen bone structures, leading to inaccurate detection and segmentation.
2. Dependency on High-Quality Images: The model accuracy depends on the quality of the X-ray images. Images with poor quality, low resolution or excessive noise can negatively impact the segmentation results.
3. Computational Resources: Training and deploying deep learning models like YOLOv8 can be computationally intensive, requiring access to powerful hardware and substantial computational resources.
4. Human Expertise: Manual annotation and preparation of accurate masks and labels require expertise in radiology and image analysis. Acquiring such skilled annotators can be resource-intensive.

# CONCLUSION

In this paper, we have explored the significant role of YOLOv8, a cutting-edge object detection and segmentation framework, in X-ray image analysis, particularly in the context of bone segmentation. The application of deep learning models, especially Convolutional Neural Networks (CNNs), has transformed the landscape of medical diagnostics, specifically in X-ray interpretation, enabling medical professionals to make more informed and accurate decisions in various scenarios. Traditional segmentation methods, while useful in many contexts, have inherent limitations when dealing with the complexity and variability of medical images, particularly in medical image segmentation tasks. Deep learning algorithms, with their ability to extract semantic information and adapt to diverse datasets, have addressed these limitations effectively. CNN-based models, such as YOLOv8, have played a significant role in advancing medical image analysis, enhancing both precision and efficiency. The combination of image segmentation and X-ray diagnosis has the potential to reshape the field of healthcare, offering automated, reliable, and detailed insights into various medical conditions.

In our research, we have focused on X-ray images of arm bones, specifically the radius, ulna, carpals, metacarpals, and phalanges. We have outlined the steps for dataset preparation, mask creation, and label generation, highlighting the format suitable for YOLOv8. The structured dataset can enable YOLOv8 to recognize and differentiate these bone structures accurately, ultimately simplifying and improving the diagnosis of X-ray images in healthcare. However, we must acknowledge existing challenges and limitations in this field. From precise manual annotation to the variability in human anatomy and the dependence on high-quality images. It is essential to recognize the need for skilled annotators, the requirement for substantial computational resources, and the potential for limitations in model generalisation.

In conclusion, YOLOv8 represents an important step in the application of deep learning to X-ray image analysis, offering a promising solution for accurate and efficient bone segmentation. While challenges and limitations persist, ongoing research and advancements in the field of medical image analysis, driven by models like YOLOv8, hold the potential to revolutionise healthcare diagnostics, clearing the path for a more advanced, effective, and meaningful medical future.